**Modern alleles in archaic human Y chromosomes support origin of modern human paternal lineages in Asia rather than Africa**

Hongyao Chen and Shi Huang*

Center for Medical Genetics, Hunan Key Laboratory of Medical Genetics, School of Life Sciences, Central South University, 110 Xiangya Road, Changsha, Hunan 410078, P.R. China

*Corresponding author: Shi Huang (huangshi@sklmg.edu.cn)

**Short title**: Archaic Y DNAs support human paternal roots in Asia

**Abstract**

Recent studies have shown that hybridization between modern and archaic humans was commonplace in the history of our species. After admixture, some individuals with admixed autosomes carried the modern Homo Sapiens uniparental DNAs, while the rest carried the archaic versions. Coevolution of admixed autosomes and uniparental DNAs is expected to cause some of the sites in modern uniparental DNAs to revert back to archaic alleles, while the opposite process would occur (from archaic to modern) in some of the sites in archaic uniparental DNAs. This type of coevolution is one of the elements that differentiate the two different models of the Y phylogenetic tree of modern humans, rooting it either in Africa or East Asia (Yuan et al., 2017). The expected reversion to archaic alleles is assumed to occur and is easily traceable in the Asia model, but is absent in the Africa model due to its infinite site assumption, which also precludes the independent or convergent mutation to modern alleles in archaic uniparental DNAs since mutations are assumed to occur randomly across a neutral genome, and convergent evolution is assumed not to occur. Here, we examined newly published high coverage Y chromosome sequencing data of two Denisovan and two Neanderthal samples to determine whether they carry modern-Homo Sapiens alleles in sites where they are not supposed to according to the Africa model. The results showed that a significant fraction of the sites that, according to the Asia model, should differentiate the original modern Y from the original archaic Y (such as those that define A00T, A0T, A1b, BT, CT, and F in the Africa model) carried modern alleles in the archaic Y samples here. Some of these modern alleles were shared among all archaic humans while others could differentiate Denisovans from Neanderthals. The observation is best accounted for by coevolution of archaic Y and admixed modern autosomes, and hence supports the Asia model, since it takes such coevolution into account.

**Introduction:**

The origin of modern humans is not yet fully understood, and questions remain open regarding the origins of our species and the processes that led it to populate the entire globe (Brauer, 1982; Henn et al., 2018; Scerri et al., 2018; Stringer and Andrews, 1988; Wolpoff et al., 1984; Wu, 2004; Yuan et al., 2017). Two competing models, termed "Multiregional" and "Recent Out-of-Africa", have long been proposed (Brauer, 1982; Stringer and Andrews, 1988; Wolpoff et al., 1984; Wu, 2004). The multiregional model considers extant people of any given region, e.g., East Asia, to be largely descended from ancient humans living in the same region at ~200-2000 ky ago, such as Peking man, with some gene flow having occurred between populations in different regions (Wu, 2004). The model has support from fossils and cultural remains, but molecular evidence has been lacking until recently (Yuan et al., 2017). Analyses based on a more complete molecular evolutionary framework, the maximum genetic diversity (MGD) theory (Huang, 2016), suggests multiregional origins for autosomes, but roots both uniparental DNAs in East Asia (Yuan et al., 2017). In this model of an Asian origin of Homo Sapiens, modernity is defined by individuals carrying modern uniparental DNAs. Migration of modern humans from East Asia to other regions would result in mating between modern and local archaic humans, and hence the gradual dilution of the autosomes carried by early modern humans from East Asia by the autosomes of the populations already inhabiting these regions before admixture/migration. Coevolution of admixed autosomes and uniparental DNAs would result in some reversion to archaic alleles in the modern uniparental DNAs carried by the admixed modern individuals, as well as some mutation to modern alleles in the archaic uniparental DNAs of admixed archaic individuals. Coevolution of uniparental DNAs and autosomes is to be expected, and is well supported by experimental studies (Gemmell and Sin, 2002; Healy and Burton, 2020; Zhu et al., 2015a). The rooting of mtDNA tree in Asia independently confirms an earlier paper (Johnson et al., 1983), and has been verified by ancient findings regarding mitochondrial DNA that shows the earlier appearance of haplogroup R compared to N, which challenges  rooting  the human mtDNA tree in Africa (Zhang and Huang, 2019a, b). The rooting of the Y tree in Asia has also been supported by findings from Y chromosomes of ancient modern humans (Chen et al., 2020a).

The Out of Africa model posits that modern humans originated in Africa, and then migrated to Eurasia, largely replacing local archaic humans with limited genetic mixing (Cann et al., 1987; Green et al., 2010; Stringer and Andrews, 1988). When the model was originally proposed, it assumed no admixture between incoming humans from Africa and the human populations inhabiting Eurasia at the time, but in recent years was revised to account for hybridization between modern and archaic humans as revealed by sequencing analyses of Neanderthals and Denisovans (Green et al., 2010; Meyer et al., 2012). The rooting of uniparental DNAs in Africa relies on the assumption that neutral mutations occur throughout the entire genome (Cann et al., 1987; Ke et al., 2001; Underhill et al., 2000), which however is known to be a poor explanatory framework for evolutionary phenomena (Huang, 2016; Kern and Hahn, 2018). The infinite site assumption emerges from the neutral framework, which states that mutations appear only once in evolutionary history, and it also leads to the related inference of derived alleles underling the Y tree topology and rooting of the Africa model (Underhill et al., 2000). However, mutation saturation and natural selection is far more common than initially thought, which would invalidate the currently accepted inference of derived alleles (Huang, 2016; Lei et al., 2018; Teitz et al., 2018; Yuan et al., 2017; Zhu et al., 2015b). The model in fact lacks self-consistency. For example, certain haplogroups contain a large number of derived alleles that define other haplogroups, e.g., A has many derived alleles for BT (42.4% of informative sites) and A and B have many derived alleles for CT (18.9% informative sites), which violates the method underlying the Africa model in the first place (Poznik et al., 2013; Yuan et al., 2017).

In contrast, haplogroups in the Asia model are defined by alleles shared by all members within a haplogroup, regardless of whether these alleles are derived or ancestral. The rooting in the Asia model of uniparental DNAs relies on the reasoning that the original haplotype should be the common type shared by most individuals, since mutations leading to alternative types should be rare events (Johnson et al., 1983; Yuan et al., 2017; Zhang and Huang, 2019a, b). The ancestor type carried by the first modern individual should have many alleles different from the outgroup to qualify as a modern type. As modern humans migrated to new places and admixed with local archaic humans, coevolution alongside admixed autosomes may have caused certain sites in modern uniparental DNAs to mutate back to archaic alleles, or certain sites in archaic uniparental DNAs to mutate into modern alleles. Such coevolution is a fundamental part of the Asia model but is not acknowledged by the Africa model given its neutral and infinite site assumptions which rule out coevolution. We here examined four previously published (two Denisovan and two Neanderthal) Y chromosome high coverage sequencing datasets to see whether these archaic Y chromosomes may carry modern alleles in sites where they are not supposed to according to the Africa model (Petr et al., 2020).

**Results and Discussion:**

We merged the genotypes of the archaic Y chromosomes with the Y-DNA haplogroup tree from the International Society of Genetic Genealogy (ISOGG, http://www.isogg.org, version 15.44, March 2, 2020). We calculated the fraction of modern alleles among informative sites for the major haplogroups, as well as for all sites combined (Table 1). Archaic Y chromosomes were found to carry a high fraction of modern alleles in sites that, according to the Asia model, differentiate the original modern Y (not yet affected by admixture) from the original archaic Y (Figure 1 and 2, Table 1), including A0000A000 (or A00T in the Africa model. A0000 is Denisovan haplotype and A000 is Neanderthal haplotype), A00 (A0T), A00A1a (A1b), A00A1b(BT), AB (CT), and ABCDE (F). For example, for sites defining A0T, the Asia model has the original modern Y chromosome carrying A0T alleles while the original archaic Y carrying A00 or non A0T alleles (Figure 2). Coevolution with admixed modern autosomes would result in mutation to modern alleles (A0T alleles) in individuals carrying the archaic Y chromosomes. Indeed, among A0T defining sites, the fraction of modern A0T alleles was 0.27, 0.31, 0.44, and 0.13 for Denisova 4, Denisova 8, Spy 94a, and Mezmaiskaya 2, respectively, which were all significantly higher than the fraction of all modern alleles among all sites/haplotypes combined (0.0019-0.0077, P<0.0001, Table 1). Other haplotypes that differ between the Asia and Africa models such as A1 and CF have too few informative SNPs to be informative in this analysis. Here, one would expect some (albeit very low) fraction of modern alleles to be found throughout the archaic Y sequences, which may represent sequencing/calling errors and background mutations. For the most part, the mutated sites were shared among the four samples (Table 2), for example, the SNP L1136 was shared among Denisova 4, Denisova 8, and Spy 94a (missing or not informative in Mezmaiskaya 2); and L1129 was shared among Denisova 8, Spy94a, and Mezmaiskaya 2 (not informative in Denisova 4). There were also mutations that differentiate the Denisovans from Neanderthals, for example, L1120 site was mutated to modern alleles in the two Denisovans but not in the two Neanderthals (Table 2). This pattern is consistent with expectation based on shared natural selection or coevolution. This is also similar to the recent finding on autosomes that Neanderthal segments recovered from modern genomes/autosomes across the world show very similar distributions along the genome/autosome, which has been interpreted based on the (unrealistic) infinite site assumption to mean a single source Neanderthal population contributing to present-day human populations (Bergstrom et al., 2020). However, given the results here and the MGD theory (Huang, 2016), sharing of alleles between archaic and modern populations in most cases is more likely a result of coevolution due to common adaptation to a shared environment or physiology, or natural selection.

**Table 1. Fraction of modern alleles in archaic Y chromosomes in sites defining modern Y haplogroups.** Genotypes of archaic Y chromosomes were merged with SNPs set of the Y-DNA haplogroup tree from ISOGG, version 15.44, March 2, 2020. The fractions of modern alleles among informative sites for the major haplogroups as well as for all sites combined were calculated and shown. For haplogroups that are defined/named differently in the Asia and Africa models, both names are listed.

| Haplogroup | Denisova 4 | | Denisova 8 | | Spy 94a | | Mezmaiskaya 2 | |
|---|---|---|---|---|---|---|---|---|
| | # Site | Fraction | # Site | Fraction | # Site | Fraction | # Site | Fraction |
| A0000A000/A00T | 41 | 0.1220 | 66 | 0.1212 | 47 | 0.0000 | 73 | 0.0685 |
| A00/A0T | 11 | 0.2727 | 16 | 0.3125 | 9 | 0.4444 | 15 | 0.1333 |
| A00A1a/A1b | 16 | 0.2500 | 17 | 0.0588 | 11 | 0.0000 | 21 | 0.0000 |
| A00A1b/BT | 140 | 0.0786 | 169 | 0.0296 | 85 | 0.1059 | 186 | 0.0000 |
| AB/CT | 67 | 0.0149 | 101 | 0.0099 | 54 | 0.0741 | 109 | 0.0000 |
| ABCDE/F | 33 | 0.0909 | 49 | 0.0612 | 30 | 0.0000 | 78 | 0.0000 |
| A00 | 746 | 0.0161 | 917 | 0.0164 | 542 | 0.0111 | 1047 | 0.0181 |
| A0 | 18 | 0.0000 | 22 | 0.0000 | 12 | 0.0000 | 21 | 0.0000 |
| B2 | 79 | 0.0000 | 109 | 0.0000 | 57 | 0.0351 | 126 | 0.0000 |
| B3 | 220 | 0.0136 | 275 | 0.0109 | 165 | 0.0242 | 285 | 0.0000 |
| G | 81 | 0.0000 | 84 | 0.0000 | 60 | 0.0167 | 103 | 0.0000 |
| I | 62 | 0.0000 | 77 | 0.0000 | 48 | 0.0000 | 92 | 0.0000 |
| H1a1 | 81 | 0.0000 | 91 | 0.0000 | 61 | 0.0000 | 120 | 0.0000 |
| E all | 3831 | 0.0068 | 4658 | 0.0052 | 2842 | 0.0067 | 5303 | 0.0008 |
| G all | 6434 | 0.0065 | 7618 | 0.0066 | 4559 | 0.0068 | 8850 | 0.0008 |
| H all | 1280 | 0.0078 | 1513 | 0.0079 | 904 | 0.0077 | 1742 | 0.0011 |
| O all | 1923 | 0.0052 | 2302 | 0.0070 | 1420 | 0.0106 | 2604 | 0.0027 |
| Q all | 2060 | 0.0063 | 2409 | 0.0062 | 1451 | 0.0048 | 2761 | 0.0011 |
| R all | 3834 | 0.0034 | 4549 | 0.0070 | 2810 | 0.0057 | 5277 | 0.0017 |
| S all | 813 | 0.0049 | 934 | 0.0043 | 549 | 0.0036 | 1077 | 0.0009 |
| All | 35920 | 0.0075 | 42702 | 0.0077 | 25764 | 0.0069 | 49320 | 0.0017 |

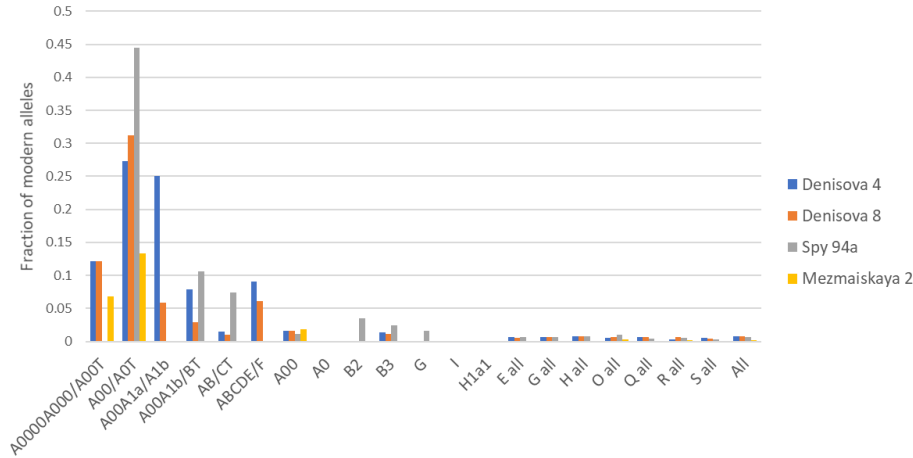

**Figure 1. Fraction of modern alleles in archaic Y chromosomes.** Fraction of modern alleles among informative sites for representative haplogroups are shown. For haplogroups whose meaning and naming are not shared by the Asia and Africa models, both names are shown for each of the concerned haplogroup.

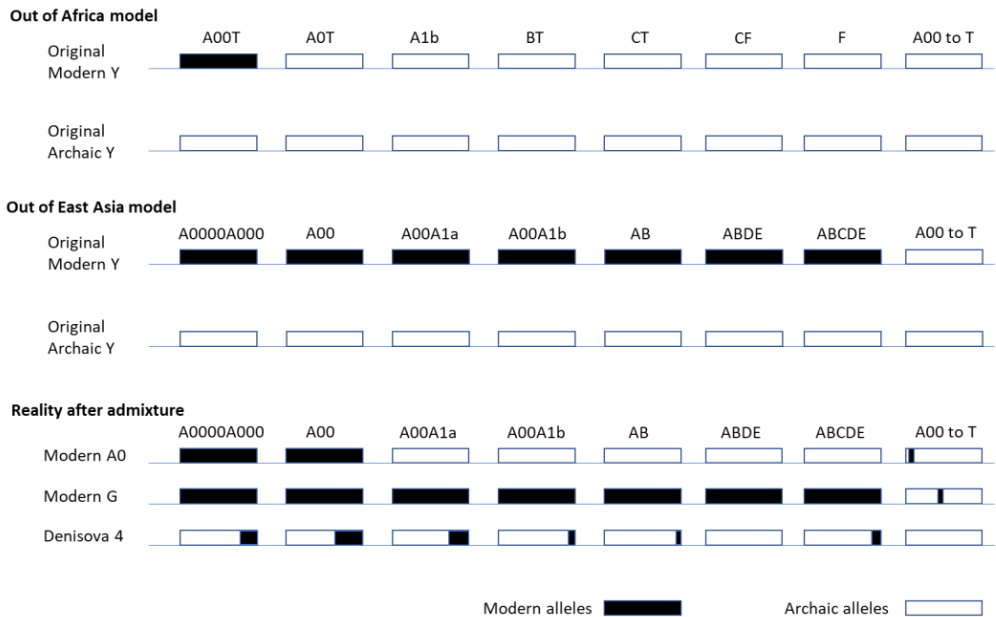

**Figure 2. Y chromosome allele patterns.** Y chromosome sequence is represented by horizontal lines and sites defining representative haplogroups are indicated by rectangular boxes (length not to scale). A0000 represents Denisovan haplotype and A000 represents Neanderthal haplotype. The term "A00 to T" represents all individual haplotypes that are defined by haplotype-specific mutations not present in archaic humans. In the case of modern A0 and G, the small black box in the "A00 to T" box indicates A0-specific mutations and G-specific mutations, respectively. A0 is a result of admixture while G is not.

**Table 2. Sharing patterns of modern alleles among archaic Y chromosomes in sites defining haplogroup A0-T.**

| SNP name | HG | Build 37 | Mutation | Denisova 4 | Denisova 8 | Spy 94a | Mezmaisk. 2 |
|----------|-----|----------|----------|-----------|-----------|---------|-------------|
| L1101 | A0-T | 6859819 | T->C | archaic | archaic | n.i. | archaic |
| L1105 | A0-T | 7590048 | C->T | archaic | archaic | n.i. | archaic |
| L1130 | A0-T | 16661010 | T->G | archaic | archaic | n.i. | archaic |
| L1155 | A0-T | 22191266 | G->C | archaic | archaic | n.i. | archaic |
| L1121 | A0-T | 14496448 | G->A | archaic | n.i. | n.i. | archaic |
| L1135 | A0-T | 18147303 | C->A | archaic | archaic | archaic | n.i. |
| L1142 | A0-T | 21288562 | C->T | archaic | archaic | archaic | n.i. |
| L1145 | A0-T | 21739790 | C->T | archaic | n.i. | n.i. | n.i. |
| L1120 | A0-T | 14496439 | G->T | **modern** | **modern** | archaic | archaic |
| L1136 | A0-T | 18914286 | A->G | **modern** | **modern** | **modern** | n.i. |
| L1143 | A0-T | 21593345 | A->G | **modern** | **modern** | n.i. | n.i. |
| L1150 | A0-T | 21903971 | A->G | n.i. | archaic | archaic | archaic |
| L1127 | A0-T | 16281810 | A->G | n.i. | n.i. | archaic | archaic |
| L1098 | A0-T | 6847637 | C->A | n.i. | archaic | n.i. | archaic |
| L1123 | A0-T | 15425676 | C->T | n.i. | archaic | n.i. | archaic |
| L1125 | A0-T | 15467768 | A->G | n.i. | archaic | n.i. | archaic |
| L1137 | A0-T | 19047091 | C->T | n.i. | archaic | n.i. | archaic |
| L1124 | A0-T | 15426344 | T->C | n.i. | n.i. | n.i. | archaic |
| L1129 | A0-T | 16596846 | T->C | n.i. | **modern** | **modern** | **modern** |
| L1132 | A0-T | 16718811 | A->G | n.i. | **modern** | n.i. | **modern** |
| L1089 | A0-T | 2887280 | G->C | n.i. | n.i. | **modern** | n.i. |
| L1095 | A0-T | 6753296 | A->G | n.i. | n.i. | **modern** | n.i. |

We also examined haplogroups where the original Y chromosomes of both archaic and modern had identical alleles. These haplogroups also happen to be no different in composition or naming between the Asia and the Africa model. Here, the archaic Y samples all had a very low fraction of carrying modern alleles ranging from 0 to 0.0078 (Figure 1 and Table 1). For example, Denisova 4 had 0 fraction of G, I, or H1a1 defining alleles, 0.0065 of G-related alleles (G all), and 0.0078 of H-related alleles (H all). Few of these haplogroups had 3 or more modern alleles and none of those few had a fraction of modern alleles greater than 0.1 in the archaic Y. Similar results were observed for the other archaic humans Denisovan 8, Spy 94a, and Mezmaiskaya 2. Also, none of these haplogroups showed higher than the expected background fraction of modern alleles across all four archaic samples.

According to the Asia model, coevolution alongside admixed modern autosomes may explain the presence of modern alleles in sites defining certain haplogroups as found here (Figure 2). At the time of admixture, modern humans had yet to evolve most haplotypes and their autosomes were largely coupled with or adapted to the original Y. Thus, sites for most modern haplotypes, except those that differentiate the modern from the archaic Y, would remain unmutated in the archaic Y chromosomes carried by the admixed archaic humans. The age for the two Neandertals, Spy 94a (38-39 ky old) and Mezmaiskaya 2 (43-45 ky old), are from direct dating of the fossil materials (Hajdinjak et al., 2018). The age for the two Denisovans, Denisova 4 (55–84 ky old) and Denisova 8 (106–136 ky old), are from molecular dating and hence less certain given the well-known uncertainty regarding the assumptions underlying the molecular dating methods (Douka et al., 2019). Their real age may be younger and close to 50 ky. Since modern humans are thought to first appear 200-50 ky ago, the ages of these four archaic humans are consistent with them having admixed modern genomes. Although future studies will be required to determine if they indeed have admixed modern autosomes, there is some evidence that introgression of modern genomes into archaic humans had occurred in the past (Chen et al., 2020b). Alternatively, random neutral mutations or natural selection may account for these modern alleles in archaic Y chromosomes. However, random neutral mutations would be expected to hit most haplogroups uniformly and in a similar manner, which is not observed. Natural selection is expected to affect different individuals of a population similarly and may find it difficult to account for the low levels of modern alleles in Mezmaiskaya 2 compared to Spy 94a (Figure 1). Thus, the results here are best accounted for by coevolution alongside admixed modern autosomes as assumed by the Asia model.

These results are difficult to reconcile with the Africa model and its assumptions. In this model, the original archaic and the original modern Y chromosomes carry the same alleles at sites that define all modern haplogroups except for the sites that define the mega-haplogroup A00-T (Figure 2). The Africa model therefore can only explain via the coevolution idea the mutation to A00-T alleles in the archaic Y as observed (Figure 1). For other alleles such as those defining haplogroups A0-T, A1b, BT, CT and F, the autosomes in the first group of modern humans from Africa were coupled with the archaic alleles of these haplogroups, and one therefore would not expect any effect of admixed modern autosomes on the archaic alleles of the archaic Y chromosomes. Also, there would be no reason for enrichment in mutation in sites for certain haplogroups (such as A00/A0T in Figure 1) but not others. Furthermore, the infinite site assumption and the neutral theory's assumption underlying the Africa model preclude any processes that would involve coevolution or adaptive mutations to explain the allele patterns in Y haplotypes. Moreover, if archaic Y already had modern alleles at some sites that define haplogroups such as BT or CT, it would invalidate the infinite site assumption's inference that mutations in such sites can only occur once and only in the BT or CT branch in the modern Y. So, the results here showing the presence of modern alleles in archaic Y chromosomes, which is best accounted for by coevolution alongside admixed modern autosomes, are inconsistent with the Africa model.

The coevolution of uniparental DNAs and admixed autosomes is assumed, and is found in the Asia, but not the Africa, model of the phylogenetic tree of uniparental DNAs. The results here on the archaic human Y chromosomes provide evidence in favor of coevolution of archaic Y chromosomes and admixed modern autosomes in admixed archaic individuals, and thus strongly support the Asia, but not Africa, model for the origin of modern human paternal lineages.

**Materials and methods:**

The vcf files of the archaic Y chromosomes were downloaded from links provided by Petr et al (2020)(Petr et al., 2020). The genotypes of the archaic samples were merged with the SNPs set of the Y-DNA haplogroup tree from ISOGG, version 15.44, March 2, 2020. The fractions of modern alleles among informative sites for the major haplogroups as well as for all sites combined were calculated. Standard statistics such as Chi squared test were performed.

**Acknowledgments:**

Supported by the National Natural Science Foundation of China grant 81171880 (S.H.).

**Additional Information:**

**Competing Interests**

The authors declare that they have no competing interests.

**Author contributions**

H.C. performed data analysis. S.H. devised the project, analyzed the data, and wrote the manuscript. All authors edited the manuscript.